\documentclass[12pt]{article}
\def\lsim{\mathrel{\rlap {\raise.5ex\hbox{$ < $}}
{\lower.5ex\hbox{$\sim$}}}}
\def\gsim{\mathrel{\rlap {\raise.5ex\hbox{$ > $}}
{\lower.5ex\hbox{$\sim$}}}}
\catcode`\@=11
\newcount\hour
\newcount\minute
\newtoks\amorpm
\hour=\time\divide\hour by60
\minute=\time{\multiply\hour by60 \global\advance\minute by-\hour}
\edef\standardtime{{\ifnum\hour<12 \global\amorpm={am}%
        \else\global\amorpm={pm}\advance\hour by-12 \fi
        \ifnum\hour=0 \hour=12 \fi
        \number\hour:\ifnum\minute<10 0\fi\number\minute\the\amorpm}}
\edef\militarytime{\number\hour:\ifnum\minute<10 0\fi\number\minute}
\def\draftlabel#1{{\@bsphack\if@filesw {\let\thepage\relax
   \xdef\@gtempa{\write\@auxout{\string
      \newlabel{#1}{{\@currentlabel}{\thepage}}}}}\@gtempa
   \if@nobreak \ifvmode\nobreak\fi\fi\fi\@esphack}
        \gdef\@eqnlabel{#1}}
\def\@eqnlabel{}
\def\@vacuum{}
\def\draftmarginnote#1{\marginpar{\raggedright\scriptsize\tt#1}}
\def\draft{\oddsidemargin -.2truein
        \def\@oddfoot{\sl preliminary draft \hfil
        \rm\thepage\hfil\sl\today\quad\militarytime}
        \let\@evenfoot\@oddfoot \overfullrule 3pt
        \let\label=\draftlabel
        \let\marginnote=\draftmarginnote
   \def\@eqnnum{(\theequation)\rlap{\kern\marginparsep\tt\@eqnlabel}%
\global\let\@eqnlabel\@vacuum}  }

\relax

\def\be{\begin{equation}}
\def\ee{\end{equation}}
\def\ba{\begin{eqnarray}}
\def\ea{\end{eqnarray}}
\def\bs{\begin{subequations}}
\def\es{\end{subequations}}
\def\d{\partial}
\def\thebibliography#1{%
\vskip 0.5cm \centerline{\bf References}
\list{%
[\arabic{enumi}]}{\settowidth\labelwidth{[#1]}
\leftmargin\labelwidth
\advance\leftmargin\labelsep
\usecounter{enumi}}
\def\newblock{\hskip .11em plus .33em minus .07em}
\sloppy\clubpenalty4000\widowpenalty4000
\sfcode`\.=1000\relax}

\renewcommand{\theequation}{\arabic{section}.\arabic{equation}}

\renewcommand{\section}{\setcounter{equation}{0}\@startsection%
{section}{1}{0mm}{-\baselineskip}{0.5\baselineskip}%
{\normalfont\normalsize\bfseries}}

\renewcommand{\subsection}{\@startsection%
{subsection}{2}{0mm}{-\baselineskip}{0.5\baselineskip}%
{\normalfont\normalsize\slshape}}

\usepackage{amscd} \usepackage{amsmath} \usepackage{amsfonts}
\usepackage{graphics}\usepackage{epic}
\usepackage{pstricks} \usepackage{latexsym}

\def\p{\psi}
\def\o{\omega}
\def\t{\tau}
\def\ch{\chi}
\def\ph{\phi}
\def\th{\theta}
\def\AD{${\rm AdS}_2$\ }
\def\SD{$S^2$\ }
\def\AT{${\rm AdS}_3$\ }
\def\ST{$ S^3$\ }
\def\AAA{${\rm AdS}_3\times S^3$\ }
\def\AA{${\rm AdS}_2\times S^2$\ }
\def\fA{\delta A}
\def\fp{\delta \psi}
\def\fch{\delta \chi}
\def\fF{\delta F}
\def\s{\sin\theta}
\def\c{\cosh\omega}
\def\arcsh{\,{\rm arcsinh}\,}
\def\ss{\sin^2\!\theta}
\def\cc{\cosh^2\!\omega}
\def\LA{\Box_{\rm AdS_2}}
\def\LS{\Box_{S^2}}
\def\GA{SL_2(\mathbb{R})}
\def\GS{SU_2(\mathbb{C})}
\def\gs{su_2(\mathbb{C})}
\def\ga{s\ell_2(\mathbb{R})}
\def\Z2{\mathbb{Z}}
\def\om{\, ^{\omega}\!}
\def\Tr{\, {\rm Tr }\, }
\def\e{\, {\rm  e } }
\begin{document}
\renewcommand{\theequation}{\arabic{section}.\arabic{equation}}

\begin{flushright}
CPTH-S023.0501 \\
\end{flushright}
\begin{centering}
%
\vskip .6cm
{\bf SOME REMARKS ON ANTI-DE SITTER D-BRANES} \\
\vspace{5pt}
\vskip .6cm
{P.M. PETROPOULOS\footnote{marios@cpht.polytechnique.fr} and
  S. RIBAULT\footnote{ribault@cpht.polytechnique.fr }}\\
\vspace{3pt}
{\it  Centre de Physique Th{\'e}orique, Ecole Polytechnique\footnote{Unit{\'e} mixte  du
CNRS et de  l'Ecole Polytechnique,
UMR 7644}}\\
{\it 91128 Palaiseau, FRANCE}\\
\vskip .8cm
{\bf Abstract}\\ 
\end{centering}
\vskip .4cm
We present some preliminary investigations about the 
\AA D3-branes in  ${\rm AdS}_3\times S^3$. We analyse the quadratic 
fluctuations of Dirac--Born--Infeld action around a given 
semi-classical D-brane configuration and compare them with results obtained 
by using conformal-field-theory techniques. We finally study classical 
motions of open strings attached to those D-branes and analyse the r{\^o}le 
of the spectral flow in this context.

\vskip .8cm
\renewcommand{\thefootnote}{\arabic{footnote}}

\setcounter{section}{0}


\section{Introduction and summary}

Among the popular string backgrounds, anti-de Sitter spaces 
play an important r{\^o}le, both from phenomenological and fundamental
viewpoints. They
naturally arise in Randall--Sundrum ``compactifications'' as well as in 
near-horizon  geometries of certain stringy black holes. The case of
\AT is even more intriguing since together with a Neveu--Schwarz 
three-form, it provides an exact string background, which is described in 
terms of the $\GA$ Wess--Zumino--Witten model \cite{su1,su11}. The latter
can be embedded in a more general exact conformal field theory (CFT),
$\GA \times \GS \times U(1)^4$. It is then automatically promoted to 
a critical type IIB superstring \cite{abs,hor}, which preserves 16 of 
the 32 type II supercharges, and has ${\rm AdS}_3\times S^3 \times T^4$  
as target space. This setting is precisely the 
near-horizon geometry of a black string,
constructed out of $Q_5$ NS five-branes and $Q_1$ fundamental strings  
of  type-IIB  theory. 
Other three-dimensional anti-de Sitter black holes can be further obtained 
as orbifolds of the above CFT. 

Despite many attempts, the resolution of the 
$\GA$ WZW model is notoriously difficult, because the group manifold is
neither compact, nor asymptotically flat. Progress on the perturbative 
closed-string spectrum, in particular, was made only recently
\cite{mo,mo1}.
This exercise remains, however, 
challenging since, among others, it provides a unique setting in
which to analyse the AdS/CFT correspondence 
beyond the supergravity approximation. Besides various phenomenological
interests for studying the D-brane configurations in the 
$\GA$ WZW model, the knowledge of their fluctuations is 
a valuable information for the string spectrum itself. 
Investigating this issue is one motivation of the present letter.

For compact Lie groups, the WZW  D-branes are rather well understood 
\cite{klim,as,fffs,sta,ars,bds,paw,ars2} 
both from the CFT and from the geometric, target-space 
viewpoints.
In fact, it has been shown in Ref. \cite{bds} that 
the semi-classical analysis of D-branes for the group $\GS$
gives {\it exact} results for numerous CFT data. This is expected to work 
for more general groups, and is likely to be related to 
some underlying supersymmetry. For the non-compact group  $\GA$,
preliminary results were obtained in Refs. \cite{sta1,sta2}.
These results were extended in \cite{bmp}, where 
  supersymmetric settings, in which the WZW  D-branes for
both  $\GA$ and $\GS$ can be embedded, were studied in detail. 
The analysis, based on classical
solutions of the low-energy (Dirac--Born--Infeld) action for the D-branes, 
reveals subtle features that are not present
in the  case of compact groups: unphysical brane trajectories, 
quantization conditions that are higher-order in the string coupling,
divergent energies etc. A purpose of this note is to go beyond the 
classical analysis of \cite{bmp}, and further investigate the spectrum of
fluctuations around the physical, supersymmetric brane solutions 
of AdS$_3 \times S^3$, namely the \AA D3-branes. 
Whether our semi-classical results are exact is also discussed in some
detail, but it remains an open question, 
which is not as easy to answer as it was in the pure
$\GS$ WZW model. 

We can summarize our results and the plan of the paper as follows.
Section \ref{rem} is a reminder of the supersymmetric D-branes of 
\AAA and of the salient properties such as the origin of their stability
related to charge quantization. In Section \ref{qflu}, 
we compute quadratic fluctuations by expanding the Dirac--Born--Infeld
(DBI)
action around a given \AA D3-brane. The spectrum, which is a combination of 
$\GS$ and (discrete) $\GA$ unitary representations, turns out to be
independent of the particular D3-brane under consideration
(i.e. independent of its electric and magnetic charges),
as expected in perturbation theory. It is nevertheless compatible with the
results obtained in Section \ref{CFT} by using a CFT analysis. 
This analysis is however incomplete because very little is known 
about $\GA$ Cardy 
boundary states and related open-string spectra. Consequently, the
CFT techniques at hand can neither exhibit charge-dependent bounds on the
permitted spins, nor identify the multiplicities of states constructed 
by spectral flow, in particular from
continuous representations (density of long-string states). 
Notice that these representations and,
more generally, the spectral-flow action are anyway
invisible at the level of the DBI dynamics. In order to get more insight on
those issues, one possibility is to turn back to the classical open
string. This is done in Section \ref{opcla}. 
After solving the type-D boundary conditions, we
describe some specific solutions, which exhibit the behaviour of short
and long strings, by using classical spectral flow to 
get the latter. The similarity with closed strings is
striking in the case of the linear (i.e. electrically-neutral)
D-brane, where all of spectral flow is preserved, whereas only half of it 
survives in the case of generic branes. This is linked to the 
reflection symmetry with respect
to this D-brane, which also hints towards its open-string spectrum, and
leads to an heuristic proof that the latter should be exactly ``half'' of the
closed-string one \footnote{Related 
considerations can be found in \cite{fest}}.

\boldmath
\section{The \AA D-brane in \AAA\!\!: a reminder}
\label{rem}
\unboldmath

We consider a type IIB string on ${\rm AdS}_3\times S^3 \times T^4$.
This is a supersymmetric background, where
the radii of \AT and \ST are equal to each other, and it admits an exact CFT
description as a supersymmetric WZW theory in $\GA \times \GS \times U(1)^4$.
The levels of the $\GA$ and $\GS$ current algebras are respectively 
$k'$ and $k$, and are related to the radius\footnote{The central
charges are 
$c_{\rm AdS_3}=3  k' / ( k' +2)$
and
$c_{S^3}=3k /(k+2)$,
and with these conventions $k'<-2$. By using (\ref{rad}), it is clear
that those central charges add up to 6.}
as follows:
\begin{equation}
L^2=-(k' +2)\alpha'=(k+2)\alpha'.
\label{rad}
\end{equation}
Supersymmetry is expected to protect 
this equation, valid a priori in the weak-curvature (large-$k$)
limit only, when $k$ and $k+2$ are undistinguishable.

We parametrize the \AAA manifold with global coordinates such that the
metric is
\begin{align*}
ds^2_{\rm AdS_3}&=L^2
\left[d\p^2+\cosh^2\! \p\left(d\o ^2-\cc \, d\t ^2\right)\right], \\
ds^2_{S^3}&=L^2
\left[d\ch^2+\sin^2\!\ch\left(d\th ^2+\ss \, d\ph ^2\right)\right], 
\end{align*}
where $\p,\o,\t \in \mathbb{R}$, $\ch,\th \in [0,\pi]$ and $\ph\in [0,2\pi]$.
The Neveu--Schwarz two-form background reads, in a convenient gauge:
\begin{align*}
B_{\rm AdS_3}&=L^2\left(\psi + {\sinh 2 \psi \over 2} \right) 
\cosh \omega \,  d\omega  \wedge  d\t,\\
B_{S^3}&=L^2\left(\chi - {\sin 2 \chi \over 2} \right) 
\sin \theta \, d\theta  \wedge  d\phi.
\end{align*}
The physical D3-branes that we will be analysing have \AA geometry,
and solve the classical equations of motion of the DBI
action (see Refs. \cite{bds,bmp}), which reads:
\begin{equation}
S_{\rm DBI}  = \int d^{4} \! \zeta \, {\cal L}_{\rm DBI} 
\ , \ \ 
{\cal L}_{\rm DBI} =  - T \sqrt{-\det \left(
\hat g +\hat B + 2\pi \alpha^{\prime} F
\right)} ; 
\label{SDBI}
\end{equation}
here  $\hat g$ and $\hat B$ are the pull-backs of the WZW backgrounds, and
$F$ the world-volume $U(1)$ field. 
The embedding of the \AA branes under consideration in the \AAA ambient
geometry is given by $\p=\p_0 , \ \ch = \ch_0$. 
The \AD part ($ \p = \p_0$) looks like a static D-string stretching
between two antipodal points lying on the boundary of AdS$_3$; for 
$\p_0 = 0$, it is just a straight line passing through the center of 
\AT (which we call {\it linear} D-brane). The induced radii
are $\ell_{{\rm AdS}_2}= L \cosh \p_0 $ and $\ell_{S^2}= L \sin \ch _0$.
The natural world-volume coordinates are $\o , \t , \th $ and $\ph $,
and the covariantly-constant $U(1)$ field carried 
by those D3-branes, in the above gauge, takes the form:
\begin{equation}
F_0 = dA_0 =- {L^2 \over 2\pi \alpha'}\left(
\p_0\cosh\o \, d\o\wedge d\t +
\ch_0\sin\th \, d\th\wedge d\ph\right) .
\label{F0}
\end{equation}
Both factors in \AA
are generalized (twined) conjugacy classes of 
the groups $\GA$ and $\GS$, invariant under 
a ``diagonal'' $\GA \times \GS$ part of the original isometry group.
This defines a natural $\ga \times \gs$ action on the functions on this 
D3-brane, whose quadratic Casimirs are the d'Alembert operators, normalized 
to unit radii:
\begin{align*}
\LA &=
-\frac{1}{\cc}\d_\t^2+\frac{1}{\c}\d_\o\c\d_\o,\\
\LS &=-\triangle_{S^2}=
-\frac{1}{\ss }\d_\ph^2-\frac{1}{\s}\d_\th\s\d_\th.
\end{align*}
A ``mini-superspace'' analysis of (\ref{SDBI}), taking into account 
the degrees of freedom corresponding only to rigid motions of the 
\AA D3-brane in the $\p$ and $\ch$ directions, shows that the 
constants $\p_0$  and  $\ch_0$ are quantized. The origin of their
quantization is, however, different. On the one hand, for the $\GS$ 
component, $\ch_0$ is discrete as a consequence of the quantization 
of the magnetic flux through \SD \cite{bds}, which ensures the stability 
of the brane against shrinking to zero size: $\ch_0 = \pi \alpha' p /L^2$. 
On the other hand, for the $\GA$ part, one can advocate that the Wilson line
around a closed string, whose momentum is the electric charge, is cyclic; 
as was argued in \cite{bmp}, locality demands that the 
quantization be more generally valid, and hence applicable to the case 
of \AD branes: $\p_0 = \arcsh \left(q/2 \pi \alpha' p T_{\rm D}\right)$ 
($T_{\rm D}$ is the D-string tension). 
We will comment shortly on this issue
in what follows.
Notice, finally, that $\ch_0$ is bounded and so is the magnetic charge $p$, while 
$\p_0$ and consequently the electric charge $q$ are not.
\boldmath
\section{Quadratic fluctuations around \AA}
\label{qflu}
\unboldmath

Our aim is now to go further, and derive the complete spectrum of small
quadratic fluctuations around these D-brane solutions. Those fluctuations
are captured in the following degrees of freedom:
\begin{equation}
\ch=\ch_0 + \fch \ , \ \ 
\p=\p_0 + \fp \ \ {\rm and} \ \ 
A= A_0 + \delta A,
\label{fluc}
\end{equation}
where 
$F_0 = dA_0 $
is given in Eq. (\ref{F0}), while
$\fch, \ \fp $ and $\delta A$ are arbitrary functions of the four
world-volume coordinates, supplemented with some gauge condition. We ignore
the fluctuations of the brane along the $T^4$.

In plugging (\ref{fluc}) in (\ref{SDBI}) and expanding out to quadratic
order, we observe that the final expression starts with linear terms. 
This is not in contradiction with our assertion that we are expanding
around a classical solution: the quantization of electric and magnetic 
charges, discussed above, forces these terms to be set to zero.
The \SD part, which is proportional 
to $\int d\ph \, d\th\, \fF_{\ph\th}$, should
be dropped because the magnetic flux is quantized
. The \AD part is proportional to $\int d\o \, d\t \, 
\fF_{\o\t}$. But unlike $S^2$, \AD is topologically trivial so
this is bound to be zero if nothing evil lingers at infinity. We will
soon be forced to make a similar assumption in order to compute the
second-order terms.
We are allowed to impose boundary conditions such
that $\int d\o\,  d\t\, 
\fF_{\o\t}=0$ and $\fp\rightarrow 0$ thanks to 
the quantized \AD charge $q=(1/4 \pi L^2)\int_{S^2}dv_2 \frac{\partial
\mathcal{L}_{\rm DBI}}{\partial \, \partial_{\varphi} A_\t} $ 
(with $\varphi$ the angular coordinate of \AT in cylindrical
coordinates). This rules out
the constant modes of $\fp$ and $\frac{\fF_{\o\t}}{\c}$, 
which are linked by the
equations of motion as we will see later. 

We now turn to the second-order terms -- the fluctuation of the
action, which, up to an irrelevant multiplicative constant, read:  
\begin{align*} 
\delta ^{(2)}S_{\rm DBI} 
\propto &
\int d\o \, d\t \, d\th \, d\ph
\\ &\frac{\c}{\s}\left[ \ss
  (\d_\th \fp)^2+(\d _\ph \fp )^2+\ss
  (\d_\th \fch)^2+(\d _\ph \fch)^2 \right. \\ &+ \left.
2\ss(\fch)^2+(\fF_{\ph\th})^2+4\s \, \fch \, \fF_{\ph\th} \right] \\
+ &\frac{\s}{\c}\left[ \cc (\d_\o
  \fch)^2-(\d_\t\fch)^2 + \cc(\d_\o
  \fp)^2-(\d_\t\fp)^2  \right. \\ &-
\left. 2\cc(\fp)^2-(\fF_{\t\o})^2-
4\c \, \fp \, \fF_{\t\o}  \right] \\
+ &\left[\frac{\c}{\s}(\fF_{\o\ph})^2+
\c\s(\fF_{\o\th})^2-\frac{1}{\c\s}(\fF_{\t\ph})^2
-\frac{\s}{\c}(\fF_{\t\th})^2 \right].
\end{align*}
In order to obtain such a concise result, we have performed some partial
integrations. A sufficient assumption is that $\fA $ and its
derivatives
be continuous
functions on \AA that vanish at the infinity of AdS$_2$. Of course, our
expression does not depend on the position of the D-brane, i.e. on the
conserved charges $p$ and $q$; this is a general feature of (twined-)
conjugacy-class D-branes in group manifolds.

The linearized equations of motion can be derived from $\delta
^{(2)}S_{\rm DBI}$: 
\begin{eqnarray*} \label{tab}
\left[\begin{array}{cc}\begin{array}{lr}
\LA-\LS+2 & 2 \\
       2\LA & \LA-\LS \end{array} & \textbf{0} \\ \textbf{0} &
  \begin{array}{lr}   \LA-\LS-2 & -2 \\
       -2\LS & \LA-\LS \end{array}\end{array}\right]
\left[ \begin{array}{c}
\fp \\ \frac{\fF_{\o\t}}{\c} \\ \fch \\ \frac{\fF_{\th\ph}}{\s} 
\end{array}\right]=0, 
\end{eqnarray*}
\begin{eqnarray*} 
&\LA  \left(-\frac{1}{\ss }\d_\ph\fA_\ph -
  \frac{1}{\s}\d_\th\s\, \fA_\th\right)\\
 &-\LS  \left(-\frac{1}{\cc}\d_\t\fA_\t+\frac{1}{\c}\d_\o\c\, \fA_\o\right)=0;
 \label{eq}
\end{eqnarray*}
these may be supplemented by the following covariant gauge condition:
\begin{equation} \label{jauge}
\frac{1}{\cc}\d_\t\fA_\t
-\frac{1}{\c}\d_\o\c\, \fA_\o
-\frac{1}{\ss }\d_\ph\fA_\ph
-\frac{1}{\s}\d_\th\s\, \fA_\th = 0.
\end{equation}
That this condition does not leave any spurious degrees of freedom
will become clear here after, when comparing these results with the CFT 
approach.

Now, let us solve these equations, while
diagonalizing the operators $\LA$ and $\LS$. We may first
look for solutions such that 
\begin{equation}
\fch=\fA_\ph=\fA_\th=-\frac{1}{\cc}\d_\t\fA_\t+\frac{1}{\c}\d_\o\c\fA_\o=0.
\label{solA}
\end{equation}
 Then if we write $\LS=j(j+1)$ we must have $\LA=j(j-1)$ or
$\LA=(j+1)(j+2)$.
Let us be more explicit on the first possibility. Given a function 
$f$ on the brane
such that $\LS f=j(j+1)f$ and $\LA f=j(j-1)f$, 
we have a solution of
the form: 
\begin{equation*}
\fp= f\ ,\ \ \frac{\fF_{\o\t}}{\c}=(j-1)f \ ,\ \
\fA_\t=\frac{1}{j}\c\d_\o f  \ ,\ \ \fA_\o=\frac{1}{j}\frac{\d_\t f}{\c}.
\label{solA1}
\end{equation*}
Note that from the case $j=0$, we learn that the constant mode 
of $\fp$ corresponds to a
constant mode of $\frac{\fF_{\o\t}}{\c}$, 
as advertised previously. As the
latter is excluded by charge conservation, the former should also be
excluded from the spectrum. 
Similar solutions are obtained with a function $g$ on the brane
such that $\LS g=j(j+1)g$ and $\LA g=(j+1)(j+2)g$, with $j\geq 0$. 
Therefore, the full set of solutions compatible with (\ref{solA})
spans the following representations of $\GS\times\GA$:
\begin{equation}
\bigoplus_{j\in \Z2 , j\geq 1}(j,j-1)   \oplus
\bigoplus_{j\in \Z2 , j\geq 0}
 (j,j+1).
\label{solA12}
\end{equation}
Note that starting from a spin-$j$ representation of $\GS$ (with $j$ a
positive integer), our equations only allow positive-Casimir integer-spin
representations of $\GA$. These are {\it discrete\footnote{For concreteness,
we remind that general discrete lowest- (highest-)weight 
representations of $\GA$,
${\cal D}^{\pm}_j$, are spanned by 
$\{\vert j, m\rangle, \, m= \pm(j+1), \pm(j+2), \ldots\}$ and 
are unitary 
for real $j\geq -1$. Their Casimir, $j(j+1)$, is positive for positive $j$.
Continuous representations have negative Casimir,
parametrized by
$j=-1/2 +is$.}
representations}, which have a
lowest- or highest-weight state and infinite dimension. We will return to
this issue in the next section.

Instead of (\ref{solA}) one could equivalently assume that
\begin{equation*}
\fp=\fA_\o=\fA_\t=-\frac{1}{\ss }\d_\ph\fA_\ph -
\frac{1}{\s}\d_\th\s\fA_\th=0.
\label{solB}
\end{equation*}
The decomposition in representations follows the same pattern
as before, 
leading thereby again to (\ref{solA12}).

Finally, the last 
class of solutions we must consider obeys 
\begin{equation*}
\fch=\fp=\fF_{\o\t}=\fF_{\ph\th}=0.
\label{solC}
\end{equation*}
For these 
\begin{equation*}
\fA_i=\d_i h,
\label{solC1}
\end{equation*}
where $h$ is any function on the brane. Now, the gauge condition
(\ref{jauge}) reads: $\LA h=\LS h$.
This restricts the allowed
solutions to members of the $\GS\times\GA$ representations:
\begin{equation*}
\bigoplus_{j\in \Z2 , j\geq 1} (j,j).
\label{solC12}
\end{equation*}

To summarize, the $\GS\times\GA$ content of the solutions of the
first-order Dirac--Born--Infeld 
equations of motion (plus gauge condition) is
\begin{equation} \label{result} 
2(0,1)\oplus 
\bigoplus_{j\in \Z2 , j\geq 1} \big(j,2(j-1)\oplus j\oplus 2(j+1)\big). 
\end{equation}
It is worthwhile to stress that no upper bound on $j$ steams out of the
present semi-classical analysis.

\section{Open-string states: towards a CFT analysis}\label{CFT}

The alternative description of D-branes is through CFT on the world-sheet.
The set of semi-classical D3-brane configurations, AdS$_2 \times S^2$, 
considered
here should be identified with the Cardy boundary states, preserving
an $\GS \times \GA$ symmetry. As far as the spectrum of quadratic
fluctuations is concerned, it should be compared with the open-string
excitations of the corresponding Hilbert space.

Let us consider the $\GS\times\GA$ CFT\footnote{The superconformal field 
theory works the same way for our purposes.}, whose operators are the
$\GS\times\GA$ currents $J^A(z)$. 
We will write highest-weight states of this theory
according to their behaviour under the zero-mode $\GS\times\GA$
subalgebra of the $\GS\times\GA$ current algebra:
$|j,j'\rangle$, where $j$ (resp. $j'$) is a representation of $\GS$ 
(resp. $\GA$) of quadratic
Casimir $j(j+1)$ (resp. $j'(j'+1)$). To construct the other states we have
to apply modes of the $J^A$s to these. 

As already mentioned, the \AAA background is such that \AT and \ST have equal
radii, so that in the corresponding WZW conformal model
the current algebras associated to the groups $\GS$ and $\GA$ have
levels related according to Eq. (\ref{rad}).
This ensures that the mass-shell 
condition\footnote{We neglect the $T^4$ contribution to the 
conformal weights since we
have ignored the fluctuations of the D-brane in those directions.}
for the lightest (level-one)
states is just  
$ j(j+1)-j'(j'+1)=0$,
which implies $j=j'$ with our conventions.  
These open-string states are obtained by acting on highest-weight states
$|j,j\rangle$ with the operators
$J^A_{-1}$.  Thus the $\GS\times\GA$ content of this set of
states is\footnote{Note that the tensor product of 
$\GA$ representations is not as
standard as the $\GS$ case since we are dealing with  
infinite-dimensional  representations. Anyway,
 the traditional spin composition rule holds for
tensor products between one finite-dimensional representation of
spin $j$ and one highest- or lowest-weight representation of spin
greater than $j$.}
\begin{equation}
\bigoplus_{j\in \Z2 , j\geq 0}[(1,0)+(0,1)]\otimes|j,j\rangle=2|0,1\rangle\oplus \bigoplus_{j\in \Z2 , j\geq 1} 2|j,(j-1)\oplus
j\oplus(j+1)\rangle.
\label{CFTres}
\end{equation}
Now, we must impose the Virasoro
$L_{1}$ condition.  If we note Adj the adjoint
representation of the group $G$ (here $\GS\times\GA$), which is realized by
the $J ^A_{-1}$s, and Rep any
representation, then the Virasoro $L_{1}$ constraint 
is a $G$-covariant map ${\rm Adj}\otimes {\rm Rep} \rightarrow
{\rm Rep}$. Hence any irreducible sub-representation of ${\rm Adj}
\otimes {\rm Rep}$ which does not appear in Rep lies in its kernel.

Thus the Virasoro constraint eliminates one of the two $|j,j\rangle$
states and we are left with the same representations that were found
in the previous section, Eq. (\ref{result}). 

Several remarks are in order here. As we already observed in
Section \ref{qflu}, when analysing the quadratic fluctuations of the 
D3-brane, the presence of $\GS$ imposes a selection rule for the allowed 
$\GA$ states: only discrete integer- and positive-spin 
(i.e. positive-Casimir) representations appear, without any bound on the
allowed (positive) spin (see (\ref{result})). There are several reasons to
believe that the agreement with the above CFT analysis is at most partial.
Firstly, in Eq. (\ref{CFTres}), the spin $j$ is necessarily bounded, as a
consequence of the unitarity of the $\GS$ current-algebra representations:
$j\leq k/2$. This bound is expected to be even stronger, and related to the
magnetic charge of the Cardy state describing the D-brane 
 under consideration. Secondly, our CFT
analysis was minimal. We considered highest-weight representations of the
$\GA$ current algebra based on discrete representations of the zero-mode, 
global algebra. It is clear, however, that other (non-highest-weight) 
representations must be taken into account, which are obtained by 
spectral flow. These are built on both discrete and continuous
representations of $\GA$, and allow for a systematic construction of the
closed-string spectrum (short-string versus long-string states) \cite{mo}.  
 However, the spectrum of quadratic
fluctuations of the DBI action around a semi-classical D-brane
configuration is due to be blind to the effect of the spectral flow, the
latter being a symmetry of the current algebra. In that respect, Eq. 
(\ref{result}) deals only with a part of the complete open-string spectrum.
In order to understand these issues and determine the multiplicities
of representations, a more complete analysis is needed, which
should rely on the knowledge of the spectrum of Cardy boundary 
states and of the exact open-string partition function.
Those data are not available at present, and their determination is beyond 
the scope of this work\footnote{Such data are accessible via 
computation of the emission
amplitude of a closed string off the brane (cylinder diagram). Among
various difficulties, we expect severe divergence problems due to 
the \AT geometry.}. Instead, studying classical motions of an open
string attached to a D-brane, can provide helpful information in that 
direction and clarify, in particular, the r{\^o}le of the spectral flow.

\boldmath
\section{Classical open strings in \AT} \label{opcla}
\unboldmath

In the following, we restrict our attention to open strings in \AT 
ending on an \AD D-brane and ignore the \SD factor (we also ignore factors of 
$k'$ and $\alpha '$, which have no influence for our purpose).
The world-sheet dynamics is described by the WZW action with 
appropriate boundary conditions. 
Let $x^\pm=\sigma^0\pm\sigma^1$ be world-sheet coordinates
($\sigma^1 \in [0,\pi]$)
and $g(x^\pm)$ the embedding of the bosonic open string in $\GA$. Let
$J_+=-\d_+gg^{-1}$ and $J_-= g^{-1}\d_-g$ be the world-sheet currents.
WZW equations of motion thus read: $\d_\mp J_\pm=0$. 
Let $\o$ be the outer automorphism of the group $\GA$, 
i.e. the conjugation
by the matrix $\Omega = \left(\begin{array}{cc} 0 & 1 \\ 1 & 0
\end{array}\right)$ (we note $\om h=\Omega h \Omega ^{-1}$ its action on
any matrix $h$ of $\GA$ or $\ga$). 
The four type-D (type-N) untwisted (twisted)
boundary conditions (see for example \cite{sta}) are
$ J_-=(-)\, ^{(\omega)}\!  J_+$.

We will first determine which
of these four possible boundary conditions are physical. 
 We consider the tangent vector to the
world-sheet along the $\sigma^0$ direction. Its norm reads:
\[ 
N=g_{\mu\nu}\d_{\sigma^0} X^\mu\d_{\sigma^0} X^\nu=
 \Tr \left(g^{-1}\d_{\sigma^0} g\right)^2
= \frac{1}{4}\Tr \left(g^{-1}J_-g-J_+\right)^2.
\]
The Virasoro constraint $T_{++}=\Tr J_+^2<0$ 
(we assume some internal unitary CFT giving a positive contribution
to the stress tensor) and the boundary conditions
enable us to determine the sign of $N$. More precisely, 
in the cases of untwisted type-N and twisted type-D
(\AD D-brane), we have $N<0$ (not only on the
boundary). Therefore, the \AD D-brane is 
physical.
On the contrary,
the twisted type-N and untwisted type-D (dS$_2$ D-brane)
do not forbid $N>0$, and lead to $g_{\mu\nu}\d_{\sigma^1}
X^\mu\d_{\sigma^1} X^\nu <0$, so they are unphysical. 
This has already been
argued in Ref. \cite{bmp} by observing that the electric field is
supercritical on the dS$_2$ D-brane. The contact with the present
discussion is made by noting that the electric field 
is related to the boundary conditions (see for
instance \cite{bp}), which can be written in terms of 
the tangent vector to the string at its end-points. 

Let us focus now on the open-string 
solutions in the case of our D-brane. The solutions of
the bulk equations are 
$g=a  (x^+)\, b(x^-)$, with $a(0)=1$ to ensure the
uniqueness of $a$ and $b$. We must impose boundary conditions,
which are equivalent 
to $\d_{\sigma^0} (b \om a)=0$ on the boundary (the type-N
boundary conditions cannot be solved so easily). Let us
introduce
\begin{eqnarray}
m &=& b \om a\ (\sigma^1=0) = b(\sigma^0) \om a(\sigma^0),
\label{m} \\
\bar{m} &=& b \om a\ (\sigma^1=\pi) = b(\sigma^0-\pi) \om a(\sigma^0+\pi).
\label{mbar}
\end{eqnarray}
These define the twined conjugacy classes to which the $\sigma^1=0$
and $\sigma^1=\pi$ ends of the string belong respectively, and allow
us to write $b$ in terms of $a$. 
Then the solution takes the form:
\begin{equation}
g=a(x^+)\, m\, ( \om a(x^-))^{-1}, 
\label{sol}
\end{equation}
where the $\bar{m}$ condition (\ref{mbar}) remains to be implemented.

This expression holds actually in a general group manifold, 
but is not very visual even in the
simple case of our \AD stretched
static D-string in AdS$_3$. We will thus restrict our 
attention to some special 
solutions such that
$J_+$ is a constant. This is a generalization of the procedure of
\cite{mo} for constructing non-trivial closed-string 
solutions by spectral-flowing
constant-$J_+$ ones (which are in that case geodesics viewed as
degenerate strings). Equation (\ref{sol}) now reads:
\begin{equation}
g=\e^{ x^+C }m\e^{-x^-\om C}, 
\label{solcon}
\end{equation}
with $C$ an $s\ell(2)$ matrix such that $\Tr \left( \e^{2\pi C}-1\right) m
\, \Omega =0$
(this condition means that $m$ and $\bar{m}$ belong to the same
twined conjugacy class).

If $\det C>0$ we have various sights according to the value of
$r=\sqrt{\det C}$. In all cases the solutions are time-like 
and stay at finite distance from the origin; $\t$ is a monotonous
function of $\sigma^0$ and the other coordinates are periodic: the
solutions should
be considered as {\it short strings}. Both ends
of the string go from $\t=-\infty$ to $\t=+\infty$ when $\sigma^0$
varies in the same range\footnote{To interpret the $\GA$ matrix $g$ of
(\ref{solcon}) as an element of AdS$_3$, we have to decompactify $\tau$.}. 
At a given world-sheet time $\sigma^0$, we see
the string turn with an angle roughly equal to $2\pi r$ around the center
of AdS$_3$; the integer part of $r$ counts the number of
complete revolutions around this point. 

If $\det C=0$, the string world-sheet degenerates into
a light-like geodesic on the D-brane.  
This geodesic goes from one boundary of \AD
to the other in finite time $\t$. At a given $\sigma^0$ the string
spans an segment of the geodesic (a point if $\psi_0=0$).

If $\det C<0$ the solution (\ref{solcon}) is spurious: the tangent vector 
$\partial_{\sigma^0}$ is space-like. This is easier to visualize in
the special case of the linear D-brane ($\psi_0 =0$), where the open string
world-sheet degenerates into a line which is a 
space-like geodesic. In the general case the string moves from a
point of one boundary (say $\o=-\infty$) 
of \AD when $\sigma^0 =-\infty$ to a point of the
other boundary when $\sigma^0 =+\infty$. It remains on the side of the
D-brane where the center of AdS$_3$ is.

The spurious character of the last solution is due to the positivity of
its $T_{++}$. 
To remedy this, we are led to make use of the spectral-flow symmetry that
we mentioned earlier. 
This symmetry of the $\GA$ WZW-model holds for open-string solutions in
the following way: 
take the general open-string solution in \AT with twisted type-D boundary
conditions and both ends of the string on the same D-brane $\psi =\psi_0$. 
Such a solution is defined by $a(x)$ and $m$, Eq. (\ref{sol}). 
Then another solution is
$$\tilde{a}(x)=\exp
\left(\frac{i}{2}wx\sigma_2\right) a(x),$$ 
with $\sigma_2$ the standard Pauli
matrix. The points $\sigma^1=\pi$ will remain on
the same D-brane if $w$ is an even integer. If $w$ is an odd integer
they move to the opposite D-brane at $\psi = -\psi_0$. So in general {\it half of the
spectral-flow symmetry is preserved by the D-brane.} 

The spectral flow changes $T_{++}$ in the same way as in the closed-string
case. Thus it can change its sign. In the case $\det C<0$, it
results in {\it long-string-like solutions} which approach the
boundary of \AT as $\sigma^0$ and $\t $ go to infinity, with both
ends of the string coming closer one from another 
and from the boundary. When $\det C>0$, 
the transformation $r\rightarrow r+1$ can be
interpreted as non-standard spectral flow using the matrix $C/r$
instead of $i\sigma_2$ (this is allowed because $C/r$ also has
eigenvalues $i$ and $-i$ so that $\exp (2\pi C/r)=1$). 

We finally concentrate on the electrically-neutral
D-brane $\psi_0=0$, which is a straight line stretching 
between two antipodal points of the boundary of \AT and passing through 
its center.
The corresponding open-string model can be 
viewed as ``half of the closed-string one''. More 
precisely,
the twisted type-D boundary conditions are of the
familiar form
\[ \psi=0\ , \ \d_{\sigma^1}\t=0\ , \ \d_{\sigma^1} \o =0\ , \]
and any classical open-string solution ending on the D-brane $\psi_0=0$ can
be extended to give a closed-string one defined for $\sigma^1 \in
[-\pi,0]$ by:
\begin{equation}\label{sym} \psi (\sigma^1)=-\psi (-\sigma^1)\ , \ \
  \t(\sigma^1)=\t(-\sigma^1) \ , \ \ \o (\sigma^1)=\o(-\sigma^1). 
\end{equation}
Of course, any closed-string solution obeying this symmetry can also be
interpreted as an open-string one.
Thus, in this case, the spectral-flow parameter $w$ is not restricted to be
even.
This reflection symmetry should also hold at the level of the spectrum.
That is, we expect not only a matching of classical excitations but also of
CFT states of the open- and closed-string theories. More, there should also
be a matching of multiplicites of those states (not only of zero
multiplicities). The conclusion of these heuristic arguments is that
{\it the open-string spectrum of the D-brane $\psi_0=0$ is
given by exactly ``half'' of the closed-string spectrum} (i.e. the chiral CFT
spectrum that should be tensored with itself and level-matched in order to
generate the closed-string spectrum). If the latter is
as proposed in \cite{mo}, then the former will
be made of the following representations of the $\GA$ current
algebra: $\hat{\mathcal{C}}^{\alpha,w}_{-1/2+is}  , \,
\hat{\mathcal{D}}^{+,w}_{j'} $, with the same range
$j'\in \left]\frac{1}{2},\frac{k'-1}{2}\right[$ 
(and the same
weight for the continuous representations). 

\vskip 0.2cm
\centerline{\bf Acknowledgements}
\vskip 0.1cm
\noindent
We thank Costas Bachas for useful discussions and comments.

\end{document}